\documentclass{revtex4}

\usepackage{graphicx}
\begin{document}
\newcommand{\nn}{\nonumber}
\newcommand{\be}{\begin{equation}}
\newcommand{\ee}{\end{equation}}
\newcommand{\ba}{\begin{eqnarray}}
\newcommand{\ea}{\end{eqnarray}}
\newcommand{\SM}{{\sc Sm~}}
\newcommand{\SUSY}{{\sc Susy~}}
\newcommand{\MSSM}{{\sc Mssm~}}
\newcommand{\LEP}{{\sc Lep}}
\newcommand{\LHC}{{\sc Lhc}}

% You should use BibTeX and revtex.bst for references
\bibliographystyle{revtex}

% number  and your conference paper identification number on the
% title page in preprint mode. Multiple \preprint commands are allowed.
\begin{flushright}
%hep-ph//01????? \\
DESY 01-192 \\
Snowmass \verb=P1_miller_0707=
\end{flushright}

\title{Proving Higgs Bosons are scalars at a Linear Collider}
% Optional argument for running titles on pages
%\title[]{}



\author{D.~J.~Miller}
\email[]{djmiller@mail.desy.de}
%\thanks{The Author would like to thank P.~M.~Zerwas, S.~Y.~Choi and J.~Wells for useful discussions.}
\affiliation{Deutches Elektronen-Synchrotron DESY, D-22603 Hamburg, Germany}


\date{\today}

\begin{abstract}
  
  The threshold dependence and angular correlations of Higgs-strahlung
  in $e^+e^-$ collisions can be used to demonstrate the spinless nature
  of the Standard Model Higgs boson. This method, and its possible
  extention to heavy neutral MSSM Higgs bosons, is discussed.

\end{abstract}
%\pacs{}

%\maketitle must follow title, authors, abstract and \pacs
\maketitle


{\bf 1.} The next generation of colliders will most likely discover
the Higgs boson, or whatever other mechanism is responsible for
Electroweak Symmetry Breaking. In order to distinguish whether any
newly discovered particle is the expected Higgs boson or something
else not yet predicted, it is important to experimentally determine
its properties. One such task is the verification of the scalar
nature of the Higgs boson.

The isotropic decay of a Higgs boson is a necessary consequence of the
zero spin of the Higgs boson but is not sufficient to confirm its
spinless nature, since other higher spin objects may also exhibit
isotropic decays~\cite{Birgit}. While such an observation is doubtless
an interesting consistency check, it is constructive to consider other
methods. In this talk I will describe the verification of the spinless
nature of the Standard Model (SM) Higgs boson from the threshold
behaviour and angular correlations of Higgs-strahlung in $e^+e^-$
collisions, $e^+e^- \to Z H$.  I then discuss the possible extension
of this method to the heavy Higgs bosons of the Minimal
Supersymmetric Standard Model (MSSM).\\[0.1cm]

{\bf 2.} As a first step, it is useful to examine the Higgs-strahlung
cross-section and angular distributions for arbitrary spin 'Higgs
bosons' without specifying the model. (Although a boson of non-zero
spin is undoubtedly {\em not} our traditional Higgs boson, I shall, in
the interests of simplicity continue to refer to it as such, and will
continue to use the term {\em Higgs-strahlung} for its radiation off a
Z boson.)  Using only the properties of the angular momentum operator,
the helicity amplitude for the $Z^* \to ZH$ current can be written
as~\cite{helicity},
\be \langle Z(\lambda_Z) H(\lambda_H) | Z^* (m) \rangle = \frac{g_W
  M_Z}{\cos \theta_W} d_{m, \, \lambda_Z-\lambda_H}^1(\theta)
\Gamma_{\lambda_Z \lambda_H},
\label{P1_miller_0707eq1} \ee 
where $\lambda_Z, \, \lambda_H$ and $m$ are the helicities of the real
Z, Higgs and virtual Z bosons respectively, and $d^j_{m m^{\prime}
  }(\theta)$ are the usual d-functions resulting from the projection
of different angular momentum eigenstates on to one another.
$\Gamma_{\lambda_Z \lambda_H}$ are {\em reduced helicity amplitudes}
which are model dependent, and normalized to be dimensionless. If the
Higgs sector is ${\cal CP}$ conserving then the parity relation
between helicity amplitudes for a Higgs boson of spin {\cal J} and
parity ${\cal P}$ can be exploited: \be \Gamma_{\lambda_Z \lambda_H} =
(-1)^{\cal J} {\cal P} \, \Gamma_{-\lambda_Z \, -\lambda_H}.
\label{P1_miller_0707eq2} \ee In this way, the Higgs-strahlung cross-section is
given by, \be \sigma = \frac{G_F^2 M_Z^6\, (v_e^2+a_e^2)}{24 \pi s^2
  \,(1-M_Z^2/s)^2} \, \beta \left[ |\Gamma_{00}|^2 + 2 \,
  |\Gamma_{11}|^2 + 2 \, |\Gamma_{01}|^2 + 2 \, |\Gamma_{10}|^2 + 2 \,
  |\Gamma_{12}|^2 \right], \label{P1_miller_0707eq3} \ee where $v_e$ and $a_e$
are the vector and axial-vector $Z$ charges of the electron; $M_Z$ is
the $Z$-boson mass, $\sqrt{s}$ the centre-of-mass energy, and the
normalized $Z$/$H$ velocity is \mbox{$\beta=2p/\sqrt{s}$} with $p$ the
$Z$/$H$ three-momentum in the centre-of-mass frame.  The Standard
Model cross-section is recovered by inserting, $\Gamma_{00}=-E_Z/M_Z$,
$\Gamma_{10}=-1$ and setting all other helicity amplitudes to zero,
resulting in an excitation curve rising linearly with $\beta$ near the
threshold, $\sigma \sim \beta$.  It is this distinctive property of
the SM which can be exploited to verify the spinless nature of the
Higgs boson.

Similarly, the polar angle distribution of the Higgs and $Z$ bosons in
the final state can be written,
\be \frac{d\sigma}{d\cos\theta} \propto \sin^2 \theta \left
    [|\Gamma_{00}|^2+2\,|\Gamma_{11}|^2 \right] + [1+\cos^2\theta]
  \left[|\Gamma_{01}|^2+|\Gamma_{10}|^2+|\Gamma_{12}|^2 \right],
\label{P1_miller_0707eq4} \ee
Inserting the SM reduced helicity amplitudes one sees that the SM
distribution of the polar angle $\theta$ is isotropic near the
threshold.

Finally, independent information can be obtained by considering the
final-state fermion distributions in the decay $Z \to f \bar f$.
Denoting the fermion polar angle in the $Z$ rest frame with respect to
the $Z$ flight direction in the laboratory frame by $\theta_*$, the
double differential distribution in $\theta$ and $\theta_*$ is given
by.
\ba \frac{d\sigma}{d\cos\theta \, d\cos\theta_*} &\propto&
\sin^2\theta \sin^2\theta_* \, |\Gamma_{00}|^2 + \frac{1}{2}
[1+\cos^2\theta][1+\cos^2\theta_*]
\left[ |\Gamma_{10}|^2 + |\Gamma_{12}|^2 \right] \nn \\
&& + [1+\cos^2\theta] \, \sin^2\theta_* |\Gamma_{01}|^2
+\sin^2\theta \,[1+\cos^2\theta_*] |\Gamma_{11}|^2 \nn \\
&& +\frac{2\,v_ea_e}{(v_e^2+a_e^2)} \frac{2\,v_fa_f}{(v_f^2+a_f^2)} \,
2 \cos\theta \cos\theta_* \left[ |\Gamma_{10}|^2-|\Gamma_{12}|^2
\right].
\label{P1_miller_0707eq5} \ea
Note that in the SM the reduced helicity amplitudes $\Gamma_{01}$ and
$\Gamma_{11}$ are zero and consequently the $[1+\cos^2\theta]
\sin^2 \theta_*$ and $\sin^2 \theta [1+\cos^2\theta_*]$ correlations 
are absent.\\[0.1cm]

{\bf 3.} Of course, these formulae are not useful unless one has
expressions for the reduced helicity amplitudes in a particular model.
To find these, consider the most general current describing the
$Z^*ZH$ vertex,
\be {\cal J}_{\mu} = \frac{g_W M_Z}{\cos \theta_W} T_{\mu
  \alpha \beta_1 ... \beta_{\cal J}} \, \varepsilon^*(Z)^{\alpha} \,
\varepsilon^*(H)^{\beta_1 ... \beta_{\cal J}},
\label{P1_miller_0707eq6} \ee
where $\varepsilon^{\alpha}$ is the usual spin-1 polarization vector
and $\varepsilon^{\beta_1 ...  \beta_{\cal J}}$ is the spin-${\cal J}$
polarization tensor which is symmetric, traceless and orthogonal to
the 4-momentum of the Higgs boson $p_H^{\beta_i}$. The tensor $T_{\mu
  \alpha \beta_1 ... \beta_{\cal J}}$ is effectively transverse due to
the conservation of the lepton current. By choosing the most general
tensor for the appropriate spin, and comparing with Eq.(\ref{P1_miller_0707eq1}),
expressions for the reduced helicity amplitudes can be found.

For example, the most general tensor for a ${\cal J^P}=0^+$ Higgs
coupling to the Z boson is given by, \be T^{\mu \alpha}=a_1
g_{\perp}^{\mu \alpha} + a_2 k_{\perp}^{\mu}q^{\alpha}, \label{P1_miller_0707eq7} \ee where
$q=p_Z+p_H$, $k=p_Z-p_H$ and $\perp$ denotes orthogonality to $q$.
Projecting with the polarization vector, and comparing this with
Eq.(\ref{P1_miller_0707eq1}), one obtains $\Gamma_{00}=(-a_1 E_Z -\frac{1}{2}
a_2 s^{3/2} \beta^2)/M_Z$, $\Gamma_{10}=-a_1$ and all other reduced
helicity amplitudes vanish.  Clearly, the SM is restored with the
choice $a_1=1$ and $a_2=0$.

In this way, it is straightforward to write down general reduced
helicity amplitudes for theories with higher spin Higgs bosons.  In
general, the coefficients of the tensors in $T^{\mu \alpha \beta_1 ...
  \beta_{\cal J}}$ ($a_1$ and $a_2$ in the $0^+$ example above) may be
functions of $\beta$, so the full behaviour of the cross section is
still unknown. However, near threshold, where $\beta$ is small, one
may make a power series expansion of these functions, and predict the
steepest possible $\beta$ dependence of the cross-section in the threshold
region. This prediction can be examined experimentally allowing one to
rule out certain spin states.

A full analysis of all possible ${\cal J^P}$ states~\cite{miller}
reveals that every spin-parity combination yields a faster than
$\beta$ rise of the cross-section at threshold, except for the $0^+$,
$1^+$ and $2^+$ states. In particular, states of ${\cal J}<3$ and odd
parity present at least a $\beta^3$ rise of the cross-section near
threshold, and for ${\cal J} \ge 3$ the behaviour is $\sim
\beta^{2{\cal J}-3}$ and $\sim \beta^{2{\cal J}-1}$ for $(-1)^{\cal J}
{\cal P}=\pm 1$ respectively.  It is therefore a simple matter to rule
out the majority of ${\cal J^P}$ states by simply measuring the
cross-section rise at
threshold~\cite{Dova:2001sq}.\\[0.1cm]

{\bf 4.} The exceptional cases must be ruled out using extra
observations.  Generally, the requirement that the $1^+$ or $2^+$
states mimic the linear $\beta$ rise of the $0^+$ state places
restrictions on the reduced helicity amplitudes. For a $2^+$ Higgs
boson, the term in the $Z^*ZH$ coupling which provides a linear rise
($g^{\alpha \beta_1} g_{\perp}^{\mu \beta_2}+g^{\alpha \beta_2}
g_{\perp}^{\mu \beta_1}$) also contributes to all possible reduced
helicity amplitudes. Consequently all of the reduced helicity
amplitudes in this model are non-zero and would provide angular
correlations in the decay of the Z boson, Eq.(\ref{P1_miller_0707eq5}), in direct
contradiction of the SM. For example, if one then observes no
$[1+\cos^2\theta] \sin^2 \theta_*$ or no $\sin^2 \theta
[1+\cos^2\theta_*]$ correlations then a $2^+$ state is ruled out.

The $1^+$ state is even easier to experimentally disprove. Firstly,
the observation of $H \to \gamma \gamma$ decays or the formation of
Higgs bosons in photon collisions, $\gamma \gamma \to H$, rules out
all spin-1 assignments as a result of the Landau-Yang theorem. In
addition, the spin-parity relation among the reduced helicity
amplitudes,~Eq.(\ref{P1_miller_0707eq2}), implies that $\Gamma_{00}$ must
vanish (indeed, this is also true for any ${\cal J^P}$ assignment
where $(-1)^{\cal J} {\cal P}=-1$, i.e. $2^-$, $3^+$, $4^-$ etc.).
Therefore the observation of a $\sin^2\theta \, \sin^2 \theta_*$
correlation in Eq.(\ref{P1_miller_0707eq5}), as predicted by the Standard
Model, eliminates this possibility. Finally, as with the $2^+$ Higgs
state, a linear $\beta$ rise of the cross-section also requires that
$\Gamma_{01}$ and $\Gamma_{11}$ be non-zero, so non-observation of
$[1+\cos^2\theta] \sin^2 \theta_*$ or $\sin^2 \theta
[1+\cos^2\theta_*]$ correlations also rules out a $1^+$ Higgs boson.

Non-zero spin Higgs bosons of non-definite parity (i.e. when the
$Z^*ZH$ vertex is parity violating) are equally straightforward to
disprove. In this case one may no longer use Eq.(\ref{P1_miller_0707eq2}) to
obtain the simple form of the (differential) cross sections seen in
Eqs.(\ref{P1_miller_0707eq3}--\ref{P1_miller_0707eq5}). In particular, the polar angle
distribution, Eq.(\ref{P1_miller_0707eq4}), is modified to include a linear
term proportional to $\cos\theta$, indicative of ${\cal CP}$
violation.  The analysis, however, proceeds as in the
fixed normality case, since the most general tensor vertex will be the
sum of the even and odd parity tensors (with appropriate phase
factors). One finds that, for every spin, only one of the even or odd
parity tensors dominates at threshold, so that the procedure outlined
above will also eliminate all mixed parity states with non-zero spins.

Also observe that this measurement can very easily rule out an odd
parity Higgs boson, which has at best a $\sim \beta^3$ rise of the
cross-section at threshold. It is, however, unable to distinguish
between the SM $0^+$ Higgs boson and a scalar Higgs boson of
indefinite parity, since their threshold behaviour
will be indistinguishable.\\[0.1cm]

{\bf 5.} The MSSM contains two Higgs doublets, and consequently five
physical Higgs states: two ${\cal CP}$ even (h and H), one ${\cal CP}$
odd (A), and two charged Higgs bosons ($H^{\pm}$). Clearly the
lightest ${\cal CP}$ even Higgs boson spin can be determined via
Higgs-strahlung, $e^+e^- \to Zh$, exactly as for the SM Higgs boson,
as described above. The heavier Higgs bosons, however, present more of
a challenge. Here I consider only the heavy neutral Higgs bosons.

At first sight our method seems hopeless for determining the spin of
the heavy Higgs bosons.  Heavy Higgs-strahlung, $e^+e^- \to ZH$, and
production of the pseudoscalar in association with the light scalar,
$e^+e^- \to Ah$, are, in this case, not useful since their
cross-sections are suppressed by the square of $\cos (\alpha - \beta)$
making them prohibitively small. The only process of use for this
method is the production of the pseudoscalar together with the heavy
scalar, $e^+e^- \to AH$. However, this process, presents two major
difficulties. Firstly, we have two unknown spins in the process
(assuming that neither spin has been experimentally determined
elsewhere), giving many more different spin combinations which must be
ruled out.  Secondly, and more significantly, in the MSSM the decay
products of the spin one virtual Z boson are scalar particles, and
consequently form a P-wave, with an expected threshold dependence
$\sim \beta^3$.  This slowly rising signal makes the threshold
measurement much more difficult for experiment.  Furthermore, any
higher spin combination which one would naturally expect to form an
S-wave, may mimic such a P-wave by having all their non-zero helicity
amplitudes $\sim \beta$ at threshold.

However, one possible exploitable feature is that, since both H and A
are scalars, the MSSM process has only one reduced helicity amplitude
$\Gamma_{00}$. Consequently the differential cross-section with
respect to the polar angle of the outgoing pair is proportional to
$\sin^2 \theta$ (see Eq.(\ref{P1_miller_0707eq4})). By contrast, when one considers
the possible tensor structures for the $Z^*HA$ vertex which produce
similar $\beta^3$ threshold cross-section dependences, one finds that
most also contribute other non-zero reduced helicity amplitudes
resulting in $[1+\cos^2 \theta]$ correlations in the differential
cross-section.

The exceptions to this rule occur when the H and A spins are identical
and non-zero. In order to discount them, one must again resort to the
further decay into fermion pairs. Note that this is much more
complicated than for the SM Higgs boson since we must now consider the
decay of a boson whose spin is unknown, in contrast to the decay of
the well studied Z boson. This is beyond the scope of this
contribution and
will be reported elsewhere~\cite{djmha}.\\[0.1cm]

{\bf 6.} In summary, the ${\cal J^P}$ quantum numbers for a SM Higgs
boson can be unambiguously determined by measuring the threshold
dependence of the Higgs-strahlung cross-section, $e^+e^- \to Z H$, and
the angular correlations \mbox{$[1+\cos^2\theta] \sin^2 \theta_*$} and
$[1+\cos^2\theta_*] \sin^2 \theta$.  The observation of a linear
dependence on $\beta$ at threshold eliminates all spin-parity
assignments different from $0^+$ except $1^+$ and $2^+$, and these
last two assignments may then be ruled out by the absence of the above
correlations. The method can be extended to the MSSM, where the
processes $e^+e^- \to Z h$ and $e^+e^- \to H A$ can be used to
experimentally verify the spinless nature of all neutral Higgs bosons
of the theory.

%\label{}

% figures should be put into the text as floats.
% Use the graphicx package (distributed with LaTeX2e).
% See the LaTeX Graphics Companion by Michel Goosens, Sebastian Rahtz,
% \begin{figure}
% \caption{}
% \label{}
% \end{figure}

% tables follow here or maybe be put in the text
% \begin{table}
% \caption{}
% \label{}
% \end{tabular}
% \end{table}

\vspace{-0.4cm}
\begin{acknowledgments}
\vspace{-0.3cm}
The Author would like to thank P.~M.~Zerwas, S.~Y.~Choi and J.~Wells for useful discussions.
\vspace{-0.3cm}
\end{acknowledgments}

%\bibliography{your bib file}

\begin{thebibliography}{99}
\bibitem{Birgit} B.~Eberle, Diploma thesis, University of Hamburg, 2001.

%\bibitem{barger} V.~Barger, K.~Cheung, A.~Djouadi, B.~A.~Kniehl and
%  P.~M.~Zerwas, Phys. Rev. {\bf D49} (1994) 79.
  
\bibitem{helicity} G.~Kramer and T.~F.~Walsh, Z. Physik {\bf 263}
  (1973) 361.

\bibitem{miller}{D.J.~Miller, S.Y.~Choi, B.~Eberle, M.M.~Muhlleitner
    and P.M.~ Zerwas, Phys. Lett. {\bf B505} (2001) 149.}

%\cite{Dova:2001sq}
\bibitem{Dova:2001sq}
M.~T.~Dova, P.~Garcia-Abia and W.~Lohmann,
%``Determination of the Higgs boson spin with a linear e+ e- collider,''
LC-PHSM-2001-055
{\it  In *2nd ECFA/DESY Study 1998-2001* 2236-2240}.

\bibitem{djmha}{D.J.~Miller, In preparation.}

\end{thebibliography}

\end{document}